\documentclass[preprint]{aastex}
\newcommand{\etal}{et al.}
\newcommand{\teff}{T$_{\rm eff}$}
\shorttitle{Stellar spectroscopy in NGC 3621}
\shortauthors{Bresolin, Kudritzki, Mendez \& Przybilla}

\begin{document}

\title{Stellar spectroscopy far beyond the Local
Group\footnotemark} \footnotetext[1]{Based on observations obtained
at the ESO Very Large Telescope.}

\author{Fabio Bresolin}
\affil{Universit\"{a}ts-Sternwarte M\"{u}nchen, Scheinerstr. 1,
81679, M\"{u}nchen, Germany; fabio@usm.uni-muenchen.de}
\author{Rolf-Peter Kudritzki}
\affil{Institute for Astronomy, 2680 Woodlawn Drive, Honolulu HI
96822; kud@ifa.hawaii.edu}
\author{Roberto H. Mendez}
\affil{Universit\"{a}ts-Sternwarte M\"{u}nchen, Scheinerstr. 1,
81679, M\"{u}nchen, Germany; rmendez@usm.uni-muenchen.de} \and
\author{Norbert Przybilla}
\affil{Universit\"{a}ts-Sternwarte M\"{u}nchen, Scheinerstr. 1,
81679, M\"{u}nchen, Germany; nob@usm.uni-muenchen.de}

\begin{abstract}
Multi-object spectroscopic observations of blue supergiants in
NGC~3621, a spiral galaxy at a distance of 6.7 Mpc, carried out with
the ESO VLT and FORS are presented. We demonstrate the feasibility of
quantitative stellar spectroscopy at distances approaching a ten-fold
increase over previous investigations by determining chemical
composition, stellar parameters, reddening, extinction and wind
properties of one of our targets, a supergiant of spectral type A1 Ia
located in the outskirts of NGC~3621. The metallicity (determined from
iron group elements) is reduced by a factor of two relative to the sun
in qualitative agreement with results from previous abundance studies
based on H~II region oxygen emission lines. Reddening and extinction
are $E(B-V) = 0.12$ and $A_V = 0.37$, respectively, mostly caused by the
galactic foreground.  Comparing stellar wind momentum and absolute
$V$ magnitude with galactic and M31 counterparts we confirm the potential
of the wind momentum-luminosity relationship as an alternative tool to
estimate extragalactic distances.
\end{abstract}


\section{Introduction}
Detailed optical work with 4m-class ground-based telescopes and the
Hubble Space Telescope on individual blue supergiants has been carried
out in recent years in the Galaxy, the Magellanic Clouds and a handful
of nearby galaxies, most notably M31 and M33
\citep{mccarthy95,kud98,monteverde97,monteverde00,venn99,venn00a,
venn00b,dufton00,smartt00}.  These stars are extremely luminous (up to
$M_V = -10$) and show signatures of strong mass outflows in their
ultraviolet, optical and infrared spectra, which are successfully
interpreted in the framework of the radiation driven wind theory
\citep{kudpuls00,kud00}. One of the main predictions of the theory,
nicely confirmed by spectral diagnostics of supergiants in the Milky
Way \citep{puls96,kud99}, Magellanic Clouds \citep{kud98,puls96},
NGC~6822 (Muschiel\-ok et al. 1999), M31 \citep{smartt00,mccarthy97}, and
M33 \citep{mccarthy95}, is a tight relationship between the modified
wind momentum (the product of mass-loss rate, terminal velocity of the
winds and square root of stellar radius) and the stellar luminosity:

\begin{equation}
\dot{M}\; v_\infty\; (R/R_\odot)^{0.5} = L^{1/\alpha}
\end{equation}

\noindent with $\alpha\simeq2/3$
\citep{kud98,kudpuls00,kud00,kud89}. This {\em Wind momentum -
Luminosity Relationship} (WLR) has a great potential as a new,
independent extragalactic distance indicator. Kudritzki \etal\/ (1999)
estimate that using the WLR technique with ten to twenty A-supergiants
per galaxy and medium resolution spectroscopy with 8m telescopes from
the ground distance moduli with an accuracy of about 0.1 mag can be
obtained out to the Virgo and Fornax clusters of galaxies. A detailed
comparison with well-known primary distance indicators is now required
in order to establish the reliability and applicability of the WLR. In
this context we have set to apply the methods of quantitative
spectroscopy to A and B supergiants in galaxies with known, mostly
Cepheid-based, distances. Among the galaxies studied within the HST
Extragalactic Distance Scale Key Project (Kennicutt, Freedman \& Mould
1995, \citealp{mould00}), NGC~3621 has been chosen for its moderate
distance (40\% that of the Virgo cluster), maximizing our chances of
detecting blue supergiants. In this Letter we discuss the
identification of suitable targets and the testing of our
spectroscopic tools with the current observational capabilities of an
8m telescope.

\section{Target selection and observations}
For the spectroscopic target selection we obtained broad-band $B$, $V$
and $I$ CCD frames of NGC~3621 with Antu, the first unit of the Very
Large Telescope (VLT) operated by the European Southern Observatory on
Cerro Paranal, equipped with FORS (FOcal Reducer and low dispersion
Spectrograph) in imaging mode. The crowded-field photometry was
carried out with the IRAF version of {\sc daophot}
(\citealp{davis94}), and indicates that the brightest stellar objects
in this galaxy appear at $V=20$, corresponding approximately to an
absolute magnitude $M_V = -10$, given the Cepheid distance modulus
$(m-M) = 29.1$ ($\pm$0.2) and average reddening $E(V-I) = 0.3$
\citep{rawson97}. We selected isolated stars brighter than $V = 22$
and with color index $-0.1< (V-I)<0.6$, corresponding to the expected
location of reddened blue supergiants in the colour--magnitude
diagram. The target selection was finalized after inspection of
archival HST WFPC2 images of a smaller field in NGC~3621 (to avoid,
where possible, contamination by neighbouring stars in our spectra),
and rejection of objects showing nebular emission on top of stellar
features, as revealed by short exposure spectra obtained during the
commissioning phase of FORS2 at the second Unit Telescope of the
VLT. Two H~II regions (one of which located within a slit together
with a stellar object) were included in the final sample, for future
comparison of chemical abundances measured independently from stellar
and nebular lines.

Optical spectra of our candidates were obtained on March 1-2,
2000, under 0.8 arcsec seeing conditions. The FORS instrument
possesses excellent multiplexing capabilities, allowing spectra
of 19 different faint objects to be taken simultaneously through
separate slitlets, each approximately 20 arcsec in length.
Fig.~\ref{galaxy} shows the distribution of the targets, and
Table~1 summarizes their position and photometric data.  Several
exposures, adding up to a total integration time of 10.7 hours,
were obtained with a single multislit setup (600 gr/mm grism,
1 arcsec slits), covering as a minimum the
3,800--4,900~\AA~wavelength range at 5\,\AA~ resolution. This
range contains most of the spectral features useful for deriving
the main stellar parameters (effective temperature, gravity and
chemical composition).

The photometric selection proved to be rather successful. We have
identified in our spectra ten supergiants, subdivided into spectral
types as follows: B (one star), A (six), F (one), and two Luminous
Blue Variable (LBV) candidates. Of the remaining targets, four have
composite spectra, while four show nebular emission superposed on
stellar features, despite our efforts to avoid such occurrences. In
Fig.~\ref{spectra} one early A-supergiant ($a$), one early
F-supergiant ($b$) and one LBV candidate ($c$) are displayed. The
quality of the extracted spectra (with a S/N ratio of order 50) is
excellent, making the classification of spectral types using galactic
standard stars, together with a first estimate of effective
temperatures, straightforward.

\section{Stellar parameters and metallicity: a case study}
Using line-blanketed model atmospheres in hydrostatic equilibrium
\citep{kurucz91,kurucz92} and detailed NLTE/LTE line formation
calculations \citep{przybilla00a,przybilla00b} a quantitative
analysis of the spectra is possible. We concentrate here on the
object at the top in Fig.~\ref{spectra} (hereafter star $a$), for
which we estimate an effective temperature
\teff\,=\,9000$\pm$400~K from the spectral classification and
subsequently derive a gravity $\log g$\,=\,1.05$\pm$0.15 (cgs)
from fitting the higher Balmer series members (from H$\gamma$
upward). A microturbulent velocity typical for luminous
supergiants, $\xi$\,=\,8~km~s$^{-1}$, is chosen for the spectrum
synthesis calculations. A more detailed analysis comprising the
remaining objects will be published elsewhere . The spectrum is
well fit with a chemical composition comparable to that of the
Large Magellanic Cloud assuming a reduction of metallicity by a
factor of two with respect to solar composition \citep[see
Fig.~\ref{slit9}]{haser98}, while a significantly lower
metallicity, as found in the Small Magellanic Cloud (0.2 solar),
can confidently be ruled out. Previous work using the emission
lines of H~II regions close to our stellar target has found a
reduction of the oxygen abundance of the same order
\citep{ryder95}. Our analysis provides additional information
concerning elements such as magnesium, iron, chromium and
titanium.

It is important to note that the spectral resolution of
5\,{\AA} allows a reasonable estimate of stellar metallicity by
synthesizing the entire observed spectrum. We have tested this
technique by reproducing the results obtained by detailed
quantitative analyses of high resolution, high S/N spectra of
Local Group A-supergiants (Przybilla, 2001, in prep.). However,
we stress that it is difficult to determine precise effective
temperatures from ionization equilibria of weak lines and
microturbulence velocities and abundances from individual lines.
Here we have to rely on the experience obtained from the analyses
of Local Group objects using higher resolution spectra. In the
next phase we intend to work at a resolution of 2\,{\AA} as
provided, for instance, by FORS 2 at the VLT.

We determined the interstellar reddening by fitting the calculated
energy distribution to the fluxes corresponding to the measured
magnitudes. A total $E(B-V) = 0.12$ is found, consistent with most, if
not all, of the effect as being due to galactic foreground reddening
\citep{burstein82}.  This value is smaller than the one obtained by
the HST Key Project $(E(B-V) = 0.23$, Rawson \etal~1997), however
their WFPC2 field is closer to the central part of the galaxy, where
dust lanes are prominent (Fig.~\ref{galaxy}). We therefore demonstrate
that, in principle, stellar spectroscopy allows us to map interstellar
extinction for accurate distance determinations.  From the Cepheid
distance modulus and an interstellar extinction $A_V = 0.37$ we derive
an absolute magnitude $M_V = -9.0$ ($\pm$0.2) for our object. The
corresponding stellar radius, calculated from the model atmosphere
flux in the $V$ filter, is $R = 250 \pm 25~R_\odot$, yielding a
luminosity $L = 3.7\times10^5~L_\odot$.

\section{Wind parameters and the WLR}
Star $a$, with its very high luminosity, is the ideal target for
testing whether it is possible to measure wind momenta of
A-supergiants in galaxies far beyond the Local Group. Mass-loss
rates $\dot{M}$ and terminal velocities $v_\infty$ are usually
determined from the shape of the H$\alpha$ stellar wind line
profile. However, because of the enormous strength of the wind
driven by the luminosity, also H$\beta$ is affected by mass-loss,
which allows a first estimate of the stellar wind momentum. As in
previous work in the Milky Way \citep{kud99} and M31
\citep{mccarthy97} we adopt "unified model atmospheres", which
allow for departures from Local Thermodynamic Equilibrium, are
spherically extended and include the hydrodynamic effects of
stellar winds in their stratification (Santolaya-Rey, Puls \&
Herrero 1997) . The calculated H$\beta$ profiles in
Fig.~\ref{wlr} show very clearly how mass-loss affects the
profile shape. The absorption profile is significantly filled by
wind emission, enabling a reasonable determination of $\dot{M}$.
Since, contrary to H$\alpha$, the H$\beta$ fit does not allow the
direct determination of terminal velocities, we adopt a typical
value $v_\infty$ = 200 km/s. Comparing with other A-supergiants
\citep{kud99,mccarthy97} the uncertainty introduced by this
assumption is of the order of 20\%. The mass-loss of the
A-supergiant obtained in this way is
$3\times10^{-6}\;M_\odot$~yr$^{-1}$, while the corresponding
modified stellar wind momentum flow is $5.9\times10^{28}$ erg
cm$^{-1}$, with an accuracy of roughly 30\%.

Fig.~\ref{wlr} compares with the WLR for galactic and M31
supergiants of solar metallicity, for which we have adopted the
stellar parameters given by Venn \etal\/ (2000), Kudritzki \etal\/
(1999) and McCarthy \etal\/ (1997), together with a more recent
distance determination to M31 equal to 783 Kpc (Holland 1998),
and a revised photometry for the M31 supergiants (Venn \etal\/
2000a).
Because of the small bolometric corrections of A-supergiants we use absolute
magnitude rather than luminosity to display the relationship.
Star $a$ is located slightly below the regression line. A
natural explanation for this is the somewhat lower metallicity
inferred from the analysis of the photospheric spectrum, which
leads to lower wind momenta
\citep{kud98,kud00,puls96,mccarthy95,haser98}. An empirical
calibration of the WLR metallicity dependence using A-supergiants
in the Magellanic Clouds is presently under way. The slight
discrepancy might also be due to the uncertainty of the H$\beta$
fit. The forthcoming VLT observations of all of our targets at
H$\alpha$ wavelengths will allow us to answer this question and
to put additional data points onto the WLR-plane. We are
confident that this will also provide an independent constraint
on the distance of NGC~3621.

\acknowledgments We thank the many people within the FORS
consortium, especially B. Muschielok and I. Appenzeller, for the
FORS2 commissioning data used in the preliminary phases of this
work.

\clearpage
\begin{deluxetable}{ccccccl}
\tablecolumns{7}
\tablewidth{0pt}
\tablecaption{FORS spectroscopic targets}

\tablehead{
\colhead{Slit no.}  &
\colhead{R.A. (J2000)}  &
\colhead{DEC (J2000)}   &
\colhead{$V$}       &
\colhead{$B-V$}     &
\colhead{$V-I$}     &
\colhead{Comments}}
\startdata
1 & 11 18 14.3 & $-$32 45 40.1 & 21.43 & 0.14 & 0.33 & \\
2 & 11 18 20.6 & $-$32 45 31.4 & 20.41 & 0.12 & 0.28 & LBV candidate\\
3 & 11 18 10.3 & $-$32 46 25.2 & 21.67 & 0.00 & 0.17 & \\
4 & 11 18 27.2 & $-$32 46 02.6 & 20.43 & 0.32 & 0.43 & star $b$\\
5 & 11 18 15.3 & $-$32 46 54.1 & 20.76 & 0.07 & 0.34 & \\
6 & 11 18 07.9 & $-$32 47 45.9 & 21.80 & 0.11 & 0.16 & \\
7 & 11 18 15.8 & $-$32 47 39.2 & 21.06 & 0.28 & 0.28 & \\
8 & 11 18 15.9 & $-$32 48 11.2 & 20.19 & 0.30 & 0.40 & \\
9 & 11 18 28.7 & $-$32 48 04.6 & 20.47 & 0.17 & 0.28 & star $a$\\
10 & 11 18 10.7 & $-$32 49 11.9 & 22.32 & 0.03 & 0.05 & \\
11 & 11 18 30.5 & $-$32 48 37.9 & 21.87 & $-$0.02 & 0.01 & H II region\\
12 & 11 18.18.1 & $-$32 49 31.8 & 20.39 & 0.25 & 0.48 & star $c$, LBV candidate\\
13 & 11 18 16.7 & $-$32 50 13.1 & 21.07 & 0.45 & 0.60 & H II region + star\\
14 & 11 18 26.6 & $-$32 50 05.8 & 21.90 & 0.20 & 0.18 & \\
15 & 11 18 20.2 & $-$32 50 35.2 & 21.68 & 0.15 & 0.31 & \\
16 & 11 18 23.3 & $-$32 51 02.7 & 21.74 & 0.07 & 0.05 & \\
17 & 11 18 27.5 & $-$32 51 05.4 & 21.01 & 0.21 & 0.33 & \\
18 & 11 18 18.2 & $-$32 51 58.9 & 21.04 & 0.00 & 0.09 & \\
19 & 11 18 19.6 & $-$32 52 10.1 & 19.51 & 0.79 & 1.05 & \\
\enddata
\end{deluxetable}

\clearpage

\begin{figure}
\caption{The 19 objects observed with the VLT and the FORS
spectrograph in NGC~3621 are marked with circles (stars) and
squares (H~II regions) on a color image obtained by combining
5-minute $B-$, $V-$ and $I-$band frames taken with the same
instrument in imaging mode. The field of view is approximately
$7\times7$ arcmin. Different coloured markers are used for clarity
only. The stars labeled as $a$, $b$ and $c$ are further described
in Fig~2. \label{galaxy}}
\end{figure}

\begin{figure}
\caption{Examples of
continuum-normalized stellar spectra obtained in NGC~3621,
displayed along with galactic supergiant stars of similar
spectral morphology and at the same resolution. Ordinate ticks
are spaced by 0.5 continuum units. (a) An early A-supergiant
(red, star $a$ in the text), bracketed by galactic A0Ia and A2Ia
template spectra (black). Line identification for the strongest
features is shown at the top (unlabeled shorter marks are used
for Fe II lines). (b) An early F-supergiant (red) compared to
galactic F0Ia and F2Ia template spectra (black). (c) A Luminous
Blue Variable (LBV) candidate. Notice the prominence of several
Fe II emission features (identified by the vertical marks), and
the presence of nebular emission, indicated by the hydrogen
Balmer lines in emission. \label{spectra}}
\end{figure}

\begin{figure}
\caption{The observed, continuum-normalized spectrum of the
A-supergiant star $a$ (in black) in the 4270-4650~\AA~wavelength
range is compared to the model predictions (\teff\,=\,9000~K,
$\log g$\,=\,1.05) at LMC (red) and SMC (blue dotted) chemical
abundances, corresponding to 0.5 and 0.2 solar, respectively. Line
identification is provided (Fe II lines are indicated by the
shorter marks). \label{slit9}}
\end{figure}

\begin{figure}
\caption{The relationship between wind momentum (in cgs units)
and luminosity (expressed in terms of absolute $V$ magnitude) for
A-supergiants in the Galaxy (open circles), M31 (filled circles)
and star $a$ (shown here by the error bars). The latter was
excluded in the calculation of the regression line. The inset
shows how the mass-loss rate was estimated by means of H$\beta$
line profile fits with NLTE unified wind model atmospheres
including stellar winds. The vertical and horizontal bars
represent 0.05 continuum units and 5\,\AA, respectively. The
observed spectrum is shown by the blue line. Our final profile
fit, corresponding to a mass-loss rate of $3\times10^{-6}
M_\odot$ yr$^{-1}$, is shown in red. Increasing the mass-loss
rate to $4\times10^{-6} M_\odot$ yr$^{-1}$ leads to the upper
dotted line, showing an incipient P-Cygni profile. Lower
mass-loss rates ($2\times10^{-6}$ and $10^{-7} M_\odot$
yr$^{-1}$) lead to the additional dotted profiles. At least part
of the discrepancies between the observed and theoretical
profiles can be attributed to the presence of metal lines (at
wavelengths indicated by the vertical marks) not accounted for in
the stellar wind line profile calculations. The deviations are
however within the observational uncertainties. \label{wlr}}
\end{figure}


\begin{thebibliography}{}
\bibitem[Burstein \& Heiles 1982]{burstein82} Burstein, D. \& Heiles, C., 1982, \aj, 87, 1165
\bibitem[Davis 1994]{davis94} Davis, L. E., 1994, A Reference
Guide to the {\sc IRAF/DAOPHOT} Package, NOAO Laboratory
\bibitem[Dufton \etal\/ 2000]{dufton00} Dufton, P. L., McErlean, N. D.,
Lennon, D. J., \& Ryans, R. S. I., 2000, \aap, 353, 311
\bibitem[Haser \etal\/ 1998]{haser98} Haser, S. M., Lennon, D. J.,
Kudrizki, R. P., Puls, J., Walborn, N. R., Bianchi, L., \&
Hutchings, J. B., 1994, \aap, 330, 285
\bibitem[Holland 1998]{holland98} Holland, S., 1998, \aj, 115,
1916
\bibitem[Kennicutt, Freedman \& Mould 1995]{kennicutt95} Kennicutt, R. C., Jr.,
Freedman, W. L. \& Mould, J. R., 1995, \aj, 110, 1476
\bibitem[Kudritzki \etal\/ 1989]{kud89} Kudritzki, R. P., Pauldrach, A. W. A., Puls, J. \& Abbott, D.
C., 1989, \aap, 219, 205
\bibitem[Kudritzki 1998]{kud98} Kudritzki, R. P., 1998, in Stellar Astrophysics for the Local Group
(eds Aparicio, A., Herrero, A. \& Sanchez, F.) 149-262 (Cambridge
University Press, Cambridge)
\bibitem[Kudritzki \etal\/ 1999]{kud99} Kudritzki, R. P., Puls, J., Lennon. D. J.,
Venn, K. A., Reetz, J., Najarro, F., McCarthy, J. K., \& Herrero,
A., 1999, \aap, 350, 970
\bibitem[Kudritzki 2000]{kud00} Kudritzki, R. P., 2000, in Unsolved problems of stellar evolution (ed
Livio, M.) 202-226 (Space Telescope Science Institute Symposium
Series No. 12, Cambridge University Press, Cambridge)
\bibitem[Kudritzki \& Puls 2000]{kudpuls00} Kudritzki, R. P. \& Puls, J., 2000, \araa, 38, 613
\bibitem[Kurucz 1991]{kurucz91} Kurucz, R. L., 1991, in Stellar atmospheres: beyond classical models
(eds Crivellari, L., \etal) 441-447 (NATO ASI Ser. C-152, Dordrecht)
\bibitem[Kurucz 1992]{kurucz92} Kurucz, R. L., 1992,  Rev. Mex. Astrofis., 23, 45
\bibitem[McCarthy \etal\/ 1995]{mccarthy95} McCarthy, J. K., Lennon, D. J., Venn, K. A., Kudritzki, R. P.,
Puls, J., \& Najarro, F., 1995, \apjl, 455, 135
\bibitem[McCarthy \etal\/ 1997]{mccarthy97} McCarthy, J. K., Kudritzki, R. P., Lennon, D. J., Venn, K.
A. \& Puls, J., 1997, \apj, 482, 757
\bibitem[Monteverde \etal\/ 1997]{monteverde97} Monteverde, M. I., Herrero, A., Lennon D. J., \& Kudritzki, R.
P., 1997, \apjl, 474, 107
\bibitem[Monteverde \etal\/ 2000]{monteverde00} Monteverde, M. I., Herrero, A. \& Lennon D. J., 2000,
\apj, in press
\bibitem[Mould \etal\/ 2000]{mould00} Mould, J. R., \etal\/, 2000,
\apj, 529, 786
\bibitem[Muschielok \etal\/ 1999]{muschielok99} Muschielok, B.,
\etal\/, 1999, \aap, 352, 40
\bibitem[Przybilla \etal\/ 2000a]{przybilla00a} Przybilla, N., Butler, K., Becker, S. R., Kudritzki, R. P. \&
Venn, K. A., 2000a, \aap, 359, 1085
\bibitem[Przybilla \etal\/ 2000b]{przybilla00b} Przybilla, N., Butler, K., Becker, S. R., \& Kudritzki, R. P.,
2000b, \aap, submitted
\bibitem[Puls \etal\/ 1996]{puls96} Puls, J., Kudritzki, R. P., Herrero, A., \etal, 1996, \aap, 305,
171
\bibitem[Rawson \etal\/ 1997]{rawson97} Rawson, D. M., \etal,
1997, \apj, 490, 517
\bibitem[Ryder 1995]{ryder95} Ryder, S. D., 1995, \apj, 444, 610
\bibitem[Santolaya-Rey, Puls \& Herrero 1997]{santolaya97} Santolaya-Rey, A. E., Puls, J. \&
Herrero, A., \aap, 323, 488
\bibitem[Smartt \etal\/ 2000]{smartt00} Smartt, S. J., Crowther,
P. A., Dufton, P. L., Lennon, D. J., Kudritzki, R. P., Herrero,
A., McCarthy, J. K., \& Bresolin, F., 2000, \mnras, submitted
\bibitem[Venn 1999]{venn99} Venn, K. A., 1999, \apj, 518, 405
\bibitem[Venn \etal\/ 2000a]{venn00a} Venn, K. A., McCarthy, J. K., Lennon, D. J., Przybilla, N., Kudritzki, R. P., \&
Lemke, M., 2000a, \apj, 541, 610
\bibitem[Venn \etal\/ 2000b]{venn00b} Venn, K. A., Lennon, D.J., Kaufer, A.,
McCarthy, J. K., Przybilla, N., Kudritzki, R. P., Lemke, M.,
Skillman, E.D. \& Smartt, S.J., 2000b, \apj, in press
\end{thebibliography}
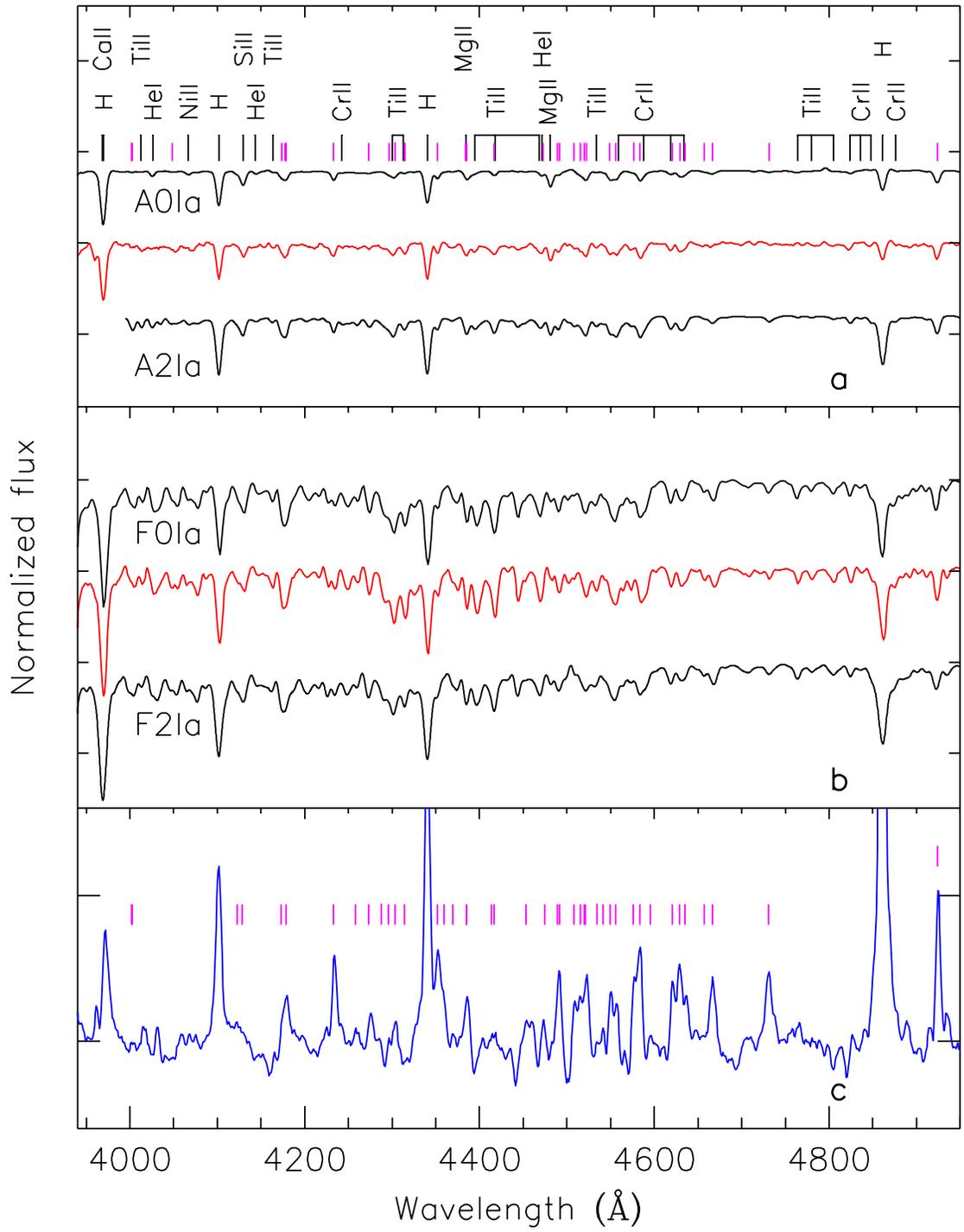
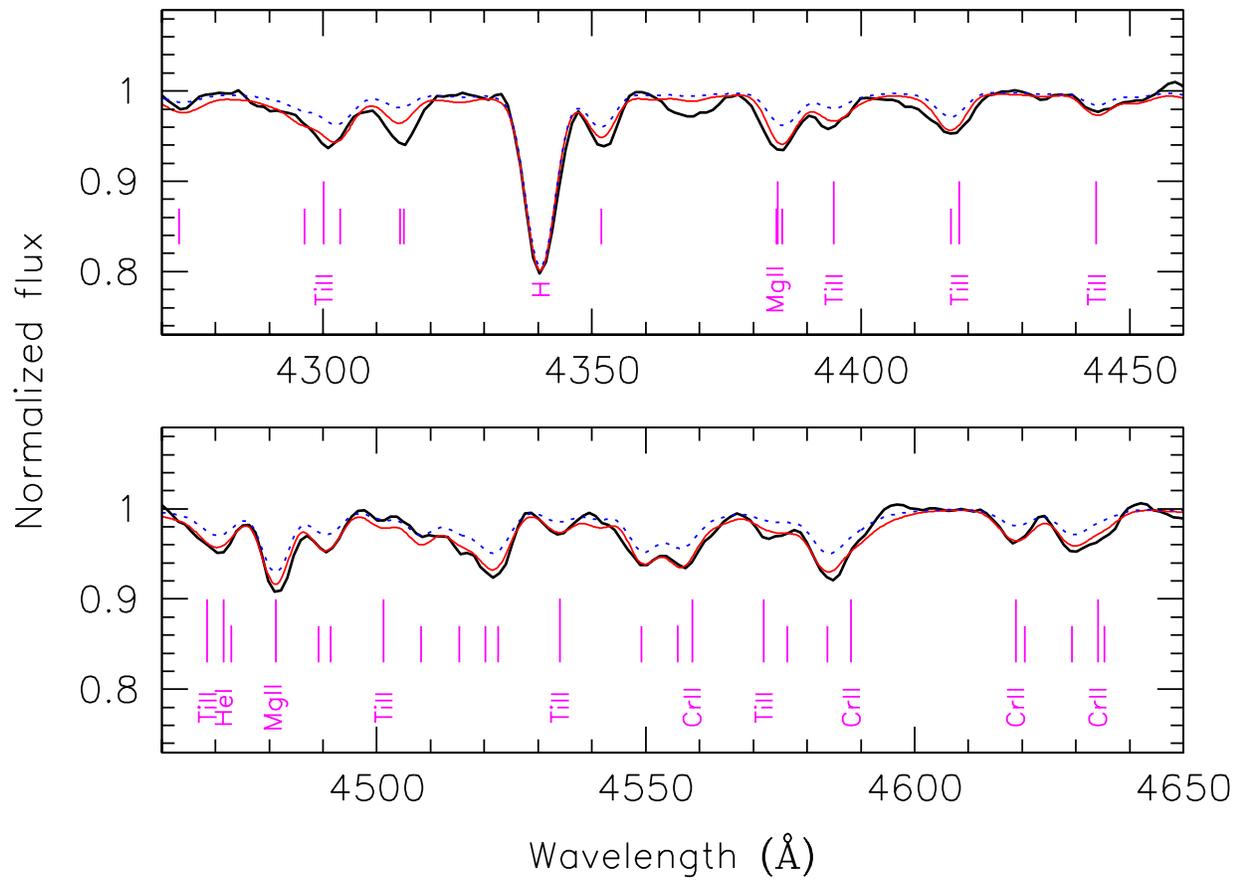
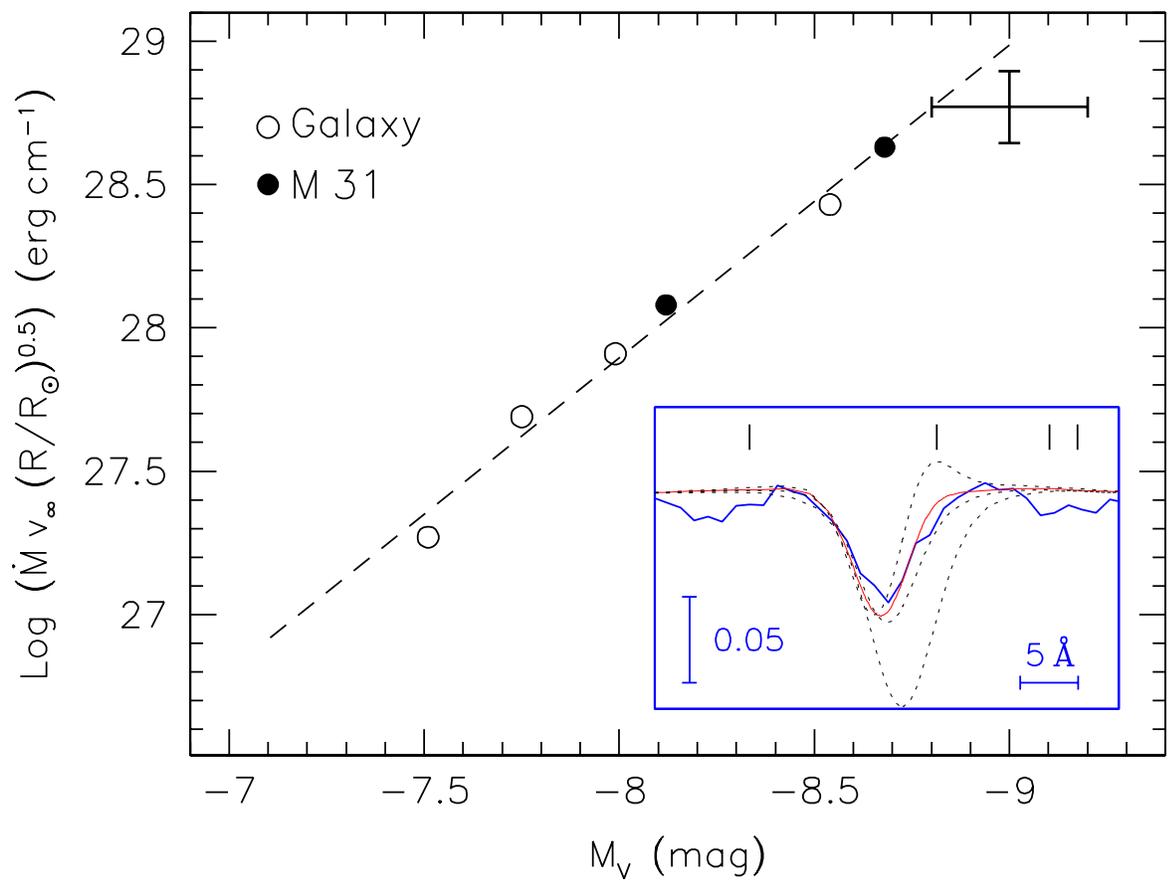
\end{document}